\RequirePackage{lineno}
\documentclass[twocolumn,prl,aps]{revtex4}
\usepackage{graphicx}
\usepackage{dcolumn}
\usepackage{bm}
\usepackage{color}
\usepackage{eso-pic}
\usepackage{amsmath}
\usepackage[retainorgcmds]{IEEEtrantools}
\usepackage{amssymb,latexsym}

\usepackage{xspace}

\newcommand{\LSMO}{$\mathrm{La}_{0.7}\mathrm{Sr}_{0.3}\mathrm{MnO}_3$\xspace}
\newcommand{\ceo}{$\mathrm{CeO}_2$\xspace}

\newcommand{\YBCO}{$\mathrm{YBa}_2\mathrm{Cu}_3\mathrm{O}_7$\xspace}
\begin{document}

\title{Ion irradiated \YBCO nano-meanders for superconducting single photon detectors }

\author{P. Amari$^{1,2}$, C. Feuillet-Palma$^{1,2}$\footnote{Electronic mail : cheryl.palma@espci.fr}, A. Jouan$^{1,2}$, F. Cou$\ddot{e}$do$^{1,2}$, N. Bourlet$^{1,2}$, E. G\'eron$^{1,2}$, M. Malnou$^{1, 2}$, L. M\'echin$^3$, A. Sharafiev$^{1,2}$, J. Lesueur$^{1,2}$ and N. Bergeal$^{1,2}$}

\affiliation{$^1$ Laboratoire de Physique et d'Etude des Mat\'eriaux, ESPCI Paris, PSL Research University, CNRS, 10 Rue Vauquelin - 75005 Paris, France}
\affiliation{$^2$ Universit\'e Pierre and Marie-Curie, Sorbonne-Universit\'e - 75005 Paris, France}
\affiliation{$^3$ Normandie Univ, UNICAEN, ENSICAEN, CNRS, GREYC, 14000 Caen, France}

\date{\today}
\begin{abstract}

We report the fabrication of few hundred microns long, hundreds of nanometers wide and 30 nm thick meanders made of \YBCO. Thin films protected by a 8 nm-thick \ceo cap layer have been patterned by high energy (a few tens of keV) oxygen ion irradiation through photoresist masks. DC and RF characterizations outline good superconducting properties of nano-meanders that could be used as Superconducting Single Photon Detectors (SSPD). By mean of a resonant method, their inductance, which mainly sets the maximum speed of these devices, has been measured on a wide range of temperature. It compares favorably to expected values calculated from the geometry of the meanders and the known London penetration depth in \YBCO.
\end{abstract}

\maketitle
Optical quantum information technologies (such as Quantum Key Distribution), low-light level depth imaging, space-to-ground communications, atmospheric remote sensing \cite{Hadfield1} require high efficiency infrared detectors with an ultimate sensitivity down to the single photon level as well as excellent time resolution. The first proof-of-concept of a single-photon detection mechanism occurring when a photon is absorbed by a superconducting nanowire biased close to its critical current \cite{Goltsman} has generated real excitement in the community in the last decade. Superconducting long, narrow and thin meanders inserted in a read-out line are the basic building blocks of this new kind of detectors called Superconducting Single Photon Detectors (SSPD).
Considerable efforts have been put forth to improve significantly their performances \cite{Hadfield} until overcoming in many ways their semiconductors based counterparts. SSPD are nowadays commercially available, based on low $T_{c}$ materials such as NbN, NbTi or WSi.

\begin{figure}[h]
\includegraphics[width=7cm]{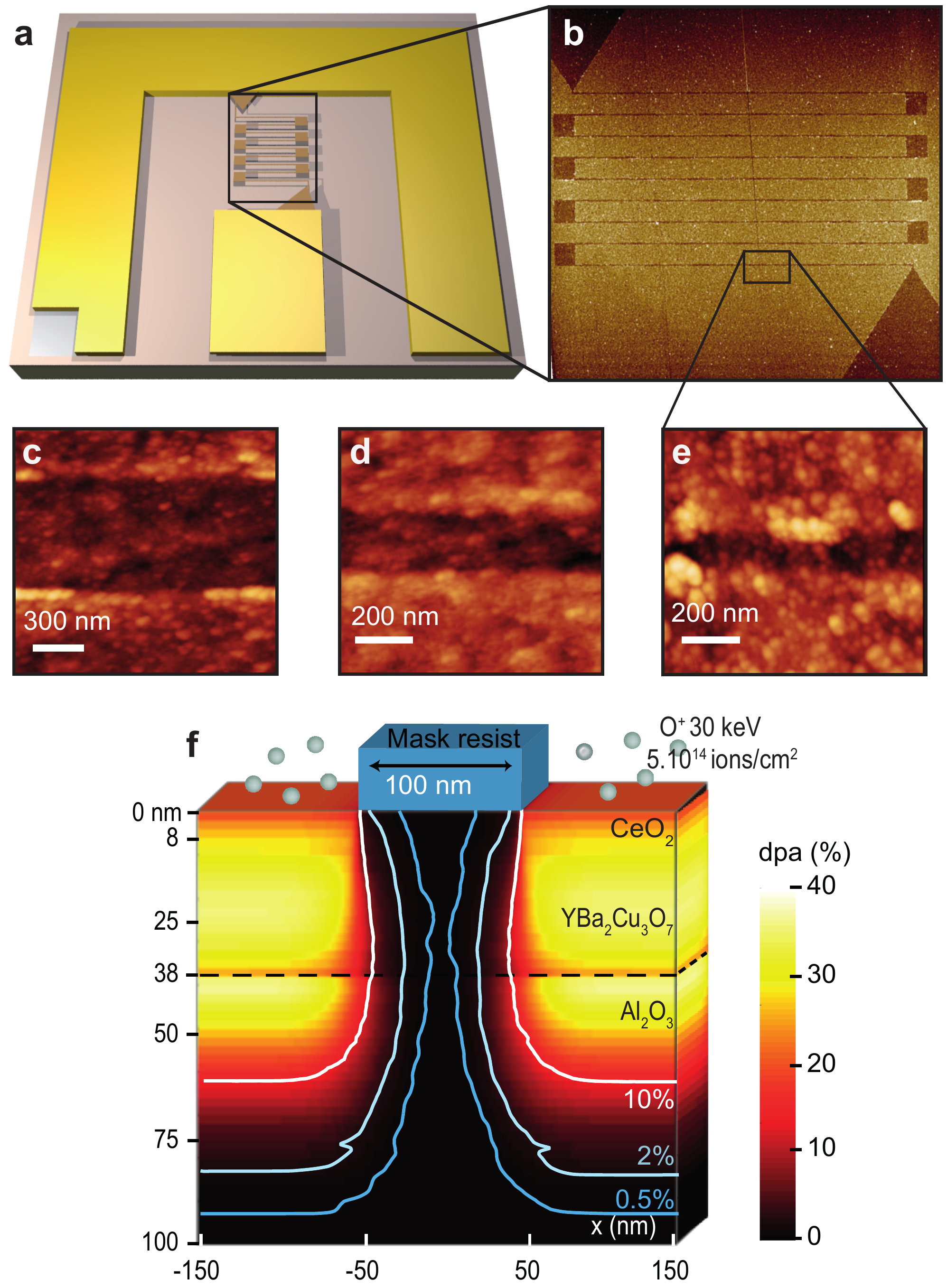}
\caption{a : Scheme of  a typical sample consisting in a sub-micrometer wide meander embedded in a gold coplanar waveguide. b-e) AFM topography images of conducting channels defined by ion irradiation (see text for explanation of the contrast). b : the whole meander,  c) focus on a 750 nm wide wire,  d) 200 nm wide and e) 100 nm. f) TRIM numerical simulation profile of the ion irradiation damages in unity of displacement per atom (dpa) produced by a 30 keV oxygen ion with a dose of $\Phi$ = 5$\times$10$^{14}$ ions/cm$^2$ for a 100 nm wide nanowire with a protecting resist (in blue). The limit of insulating \YBCO corresponding to $dpa=10 \%$ has been plotted in white. The 2\%(blue) and 0.5\% (light blue) dpa lines are also shown (see text). } 
\label{Figure1}
\end{figure}
\vskip  0.2cm

Despite tremendous advantages offered by those detectors such as high quantum efficiency ($\sim$ 93 $\%$ at around $\lambda$ $\sim$ 1.54 $\mu$m) \cite{Marsili1}, high operating frequency ($\sim $ 1 GHz \cite{Rosfjord}), low intrinsic dark count rate ($\sim$ 0.1 cps \cite{Shibata}), low jitter ($\sim$18 ps full width at half maximum \cite{You}) and broad spectral range (from visible to mid-infrared $\lambda \sim$ 10.6 $\mu$m) \cite{Marsili,Korneev}, the operating temperature well below 4K for optimal performances prevents large scale applications of these detectors. The fabrication of \YBCO based SSPD would then provide two main benefits : the ability of working at higher temperature, reducing thus drastically the cryogenic operating costs, and at higher frequency operation due to very short relaxation times \cite{Lindgren}. \YBCO implementation suffers from the main difficulty of preserving pristine superconducting properties during the nano-patterning process, because of defects creation or oxygen out-diffusion. Recently Arpaia et al. reported on the fabrication of \YBCO / \LSMO tens of microns long nanowires with a 100x50 $nm^2$ cross section \cite{ArpaiaH} of \YBCO\cite{Arpaia}, and \YBCO/Au\cite{ArpaiaI} hundred of nanometers long ones with a cross section down to 50x50 $nm^2$, while preserving superconducting properties. The longest one in a meander shape (430 $\mu m$ long) was 1 $\mu$m wide on a 12 nm thick film. Small degradations were observed, as the critical temperature was lowered down to $\sim$ 70 K only \cite{Zbinden}. However, no report of long and narrower meander on \YBCO thin films has been made up to now.

In this work, we report on a technique that allows making hundred of microns long nano-meanders with a width down to 100 nm from ion-irradiated 30 nm-thick \YBCO films. DC and RF characterizations of meander-shaped nanowires evidence good superconducting properties that are reproducible and stable in time. The inductance $L$ of the devices, a key parameter that governs the reset time $L$/R$_L$ of an SSPD ($R_L$ is the load resistance of the circuit), has been measured. Its value and its temperature dependence given by the London penetration depth one, agree very well with the expected ones in \YBCO . 

The fabrication method used here relies on the high sensitivity to disorder of the d-wave superconducting order parameter of High Temperature Superconducting (HTS) materials. Irradiation with high-energy oxygen ions accelerated under a few tens of kV, generates punctual defects in the  crystal structure that increase the normal resistivity and reduce $T_{c}$.  A superconducting-insulating transition can be induced by increasing the defect density beyond a threshold. Superconducting devices can be designed by irradiating a \YBCO thin film through a thick resist template patterned either by optical or e-beam lithography \cite{bergeal,bergealjap}. The exposed areas become insulating while the areas protected by the resist remain superconducting. The method was successfully applied to design Josephson TeraHertz Heterodyne detectors \cite{malnouapl,malnoujap,sharafiev} and Superconducting Quantum Interference Filters \cite{ouanani} made from \YBCO commercial Ceraco \cite{ceraco} thin films. 

For this study, we used 30 nm-thick YBCO commercial thin films grown on a sapphire substrate and capped with a 8nm-thick \ceo layer, processed as follows.
A 50 $\Omega$ Coplanar Waveguide (CPW) line is first designed on top of the \YBCO layer by e-beam lithography (Figure \ref{Figure1} a). The unprotected \ceo layer is removed by a 500 eV $Ar^{+}$ Ion Beam Etching (IBE) step, and an \textit{in-situ} gold layer is deposited to form the CPW line, in good electrical contact with \YBCO. Then the meander is patterned with a second e-beam lithography in a ma-N negative resist on top of the \ceo protecting layer left at the end of the CPW line. Finally, a 30 keV oxygen ion irradiation at a dose of 5 $\times$ $10^{14}$ at/$cm^2$ through the photoresist mask defines the nano-meander by insulating its edges from the rest of the CPW line, which is itself protected by the gold layer.

The main advantages of this method is that the HTS nanowire is totally encapsulated between the insulating \YBCO film on the side and the \ceo layer on the top, preventing oxygen loss. The energy and the fluence of oxygen ions necessary to make \YBCO insulating without altering the nanowire have been optimized through numerical simulation using the Monte Carlo based code TRIM \cite{SRIM}. 
This is illustrated for the 100 nm wide meander in Figure \ref{Figure1}f, by a color-plot of the calculated local displacement per atoms (dpa) which quantifies the defect density. For dpa $> 10 \%$, the material is insulating (white line). 

\begin{figure}[h]
\includegraphics[width=8.5cm]{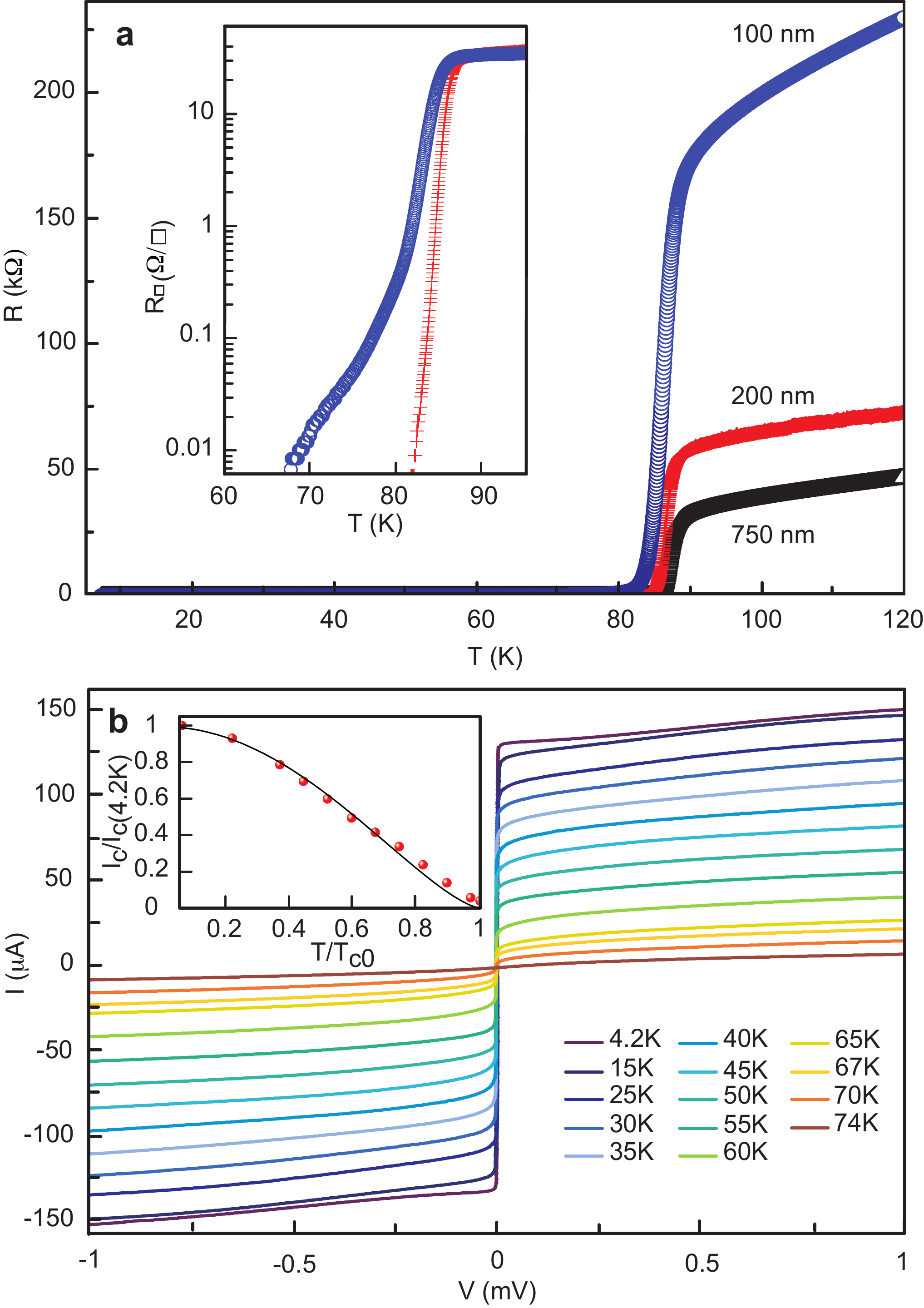}
\caption{ a : Resistance as a function of Temperature for \YBCO/\ceo nano-meanders of different widths -100 nm (blue circle), 200 nm (red), 750 nm (black)-. b : Current-voltage characteristics of a 100 nm wide nanowire at different temperatures indicated in the frame. Inset : Corresponding dependence of the critical current with temperature. The black line is a fit with the Bardeen formula (see text) } .
\label{Figure2}
\end{figure}
\vskip -0.2cm

We made wires with length ranging from 20 $\mu$m to 750 $\mu$m and width from 100 nm to 750 nm, folded into a 80$\times$80 $\mu$m$^2$ square. Each nanowire is inserted in the central part of the CPW line (Figure \ref{Figure1}a).The sample is connected via a PC-board, and mounted onto a cold sample holder in a cryogen-free cryostat. The DC Resistance versus Temperature (R-T) and the Current versus Voltage (I-V) curves are recorded through four probes measurements. RF measurements are performed through semi-rigid co-axial wires. It's worth mentioning that the devices have been thermally cycled and measured several times during more than two years, with no change in their characteristics whatsoever.  In this paper, we report specifically on three nano-meanders whose dimensions are 750 nm $\times$ 750 $\mu$m, 200 nm $\times$ 450 $\mu$m and 100 nm $\times$ 450 $\mu$m respectively. We checked that the actual widths of the conducting channels correspond to the designed ones by making Atomic Force Microscopy (AFM) images (Figure \ref{Figure1} b to e), thanks to a tiny swelling of the layer under irradiation ($\sim 1 nm$ in height). The origin of such a swelling, already reported in the literature \cite{Hurand}, is not fully established yet : it is probably related to partial amorphization of the film/substrate system.

In Figure \ref{Figure2}a, R-T curves are reported for the nano-meanders. 
The 750 nm wide nanowire has a midpoint critical temperature $T_{c}$ of 86 K (defined as a 50 $\%$ resistance drop), identical to the one of the unprocessed film. $T_{c}$ is decreased down to 83 K and 81 K for the 200 nm and the 100 nm wide ones respectively.  This can be understood in the frame of TRIM calculations, creation  of defects under irradiation, as shown in Figure \ref{Figure1}f. Indeed, ions mainly penetrate in the unprotected areas, but some are scattered, penetrate under the protected area over a mean distance of 25 nm in our case, and produce a small amount of defects (dpa). We can calculate the corresponding bulk $T_{c}$ reduction, as for instance reported in Figure \ref{Figure1}f by the light blue (dpa = 2\%) and blue (dpa = 0.5\%) lines, namely a $T_{c}$ reduction of 30\% and 10 \% respectively. The local $T_{c}$ in the actual nanostructure is more complex to evaluate, but is certainly between these values and the one for a pristine sample, and follows the trend : the narrower the wire, the lower $T_{c}$, as we measure. In addition, the lateral gradient in defect density induces a gradient in local $T_{c}$, and therefore a broader resistive transition in temperature. We observed experimentally this effect as shown in inset of Figure \ref{Figure2} a, which is a zoom of the main part for the narrower wires. The 100 nm wire R-T transition extends over $\sim15 K$, as compared to $\sim1.5 K$ only for the 200 nm wide one. In addition, a characteristic "foot" in the transition is seen in the former. Then $T_{c0}$ corresponds to the temperature  where the resistance reaches zero with the usual criteria of an electric field of $1 \mu V.cm^{-1}$ within the sample equivalent to a resistance per square of $R_{\square}\sim 10^{-2}\Omega/\square$.  $T_{c0}$ is $67\pm0.1 K$ and $81.5\pm0.1 K$ for the 100 nm and 200 nm wide wires respectively.

The I-V characteristics of the 100 nm wide nanowire is presented in Figure \ref{Figure2} b for temperatures ranging from 4.2 K to 74 K. The critical current density is $4.3 \times 10^6$ A.$cm^{-2}$ at 4.2 K. This value is comparable with the ones reported for \YBCO nanowires \cite{Rauch,Arpaia,Nawaz}, and is only two orders of magnitude smaller than the depairing limit $j_d \sim 5 \times 10^8 A.cm^{-2}$. Its temperature dependence follows the Ginzburg-Landau expression generalised by Bardeen for dirty superconductors \cite{Bardeen} over the whole range in temperature (inset \ref{Figure2} b) : $I_c(T)/I_c(4.2K) =\left(1-\left( T/T_{c0} \right)^2\right)^{-3/2}$
with $T_{c0} = 67 K$ as previously measured. 
DC measurements highlight that our technique provides a reliable way to engineer long, narrow and thin superconducting meanders in \YBCO thin films.

\begin{figure}[h]
\includegraphics[width=8.5cm]{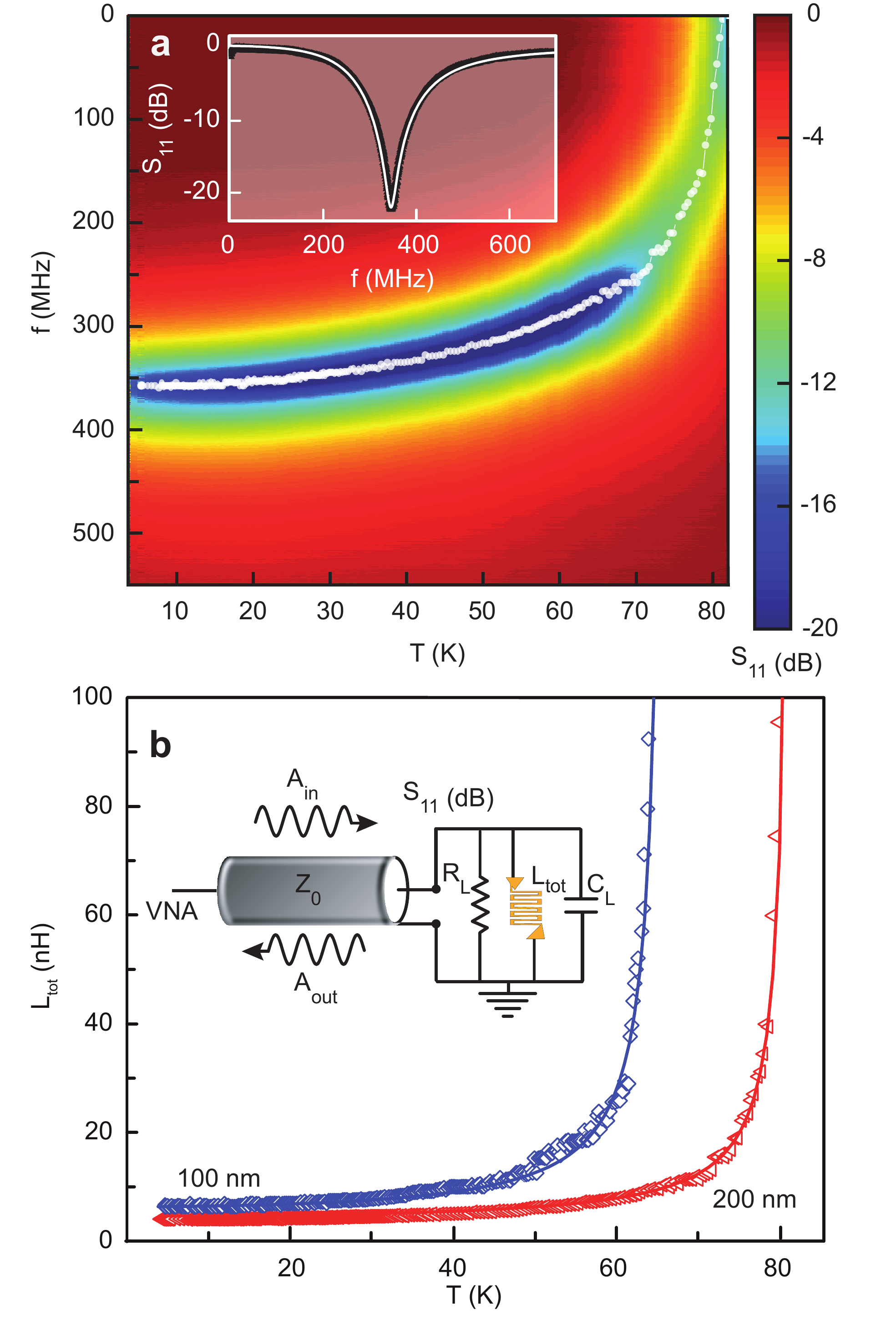}
\caption{a : Color-plot of $S_{11}$ as a function of frequency and temperature for the 200 nm wide meander showing the shift of resonant frequency $f_0$ with temperature. Inset a : $S_{11}$  at T = 30 K for the 200 nm wide meander (open squares) with the mismatch fit (solid white line), $f_{0}=348 MHz$. b : Temperature dependance of  $L_{tot}$ for the meanders : 100 nm wide (blue diamonds), 200 nm (red triangles), and the corresponding fits in continuous lines.  Inset b : SchematicÄ view of the resonant circuit.}
\label{Figure3}
\end{figure}
\vskip -0.2cm

Detection events in SSPD devices generate a voltage pulse whose characteristic times are dominated by the ratio $L$/R$_L$, where $L$ is the inductance and $R_{L}$ the load resistance of the circuit. We designed a set-up to measure the inductance of the meanders in the low GHz frequency range.
In a microwave circuit, the reflection coefficient S$_{11}$ at a discontinuity between a transmission line and a load circuit is given by : $S_{11}=(Z_L-Z_0)/(Z_L+Z_0)$,
where Z$_{0}$ = 50  $\Omega$ is the characteristic impedance of the transmission line and Z$_{L}$ is the load impedance. In this experiment, a lumped element resonant circuit involving the nano-meander is directly connected to the CPW line (inset figure \ref{Figure3} b) , and the reflection coefficient S$_{11}$ is measured using a Vector Network Analyser (VNA). The device is connected in parallel with a 24 pF Surface Mounted Device (SMD) capacitor $C_L$ and a SMD resistor $R_L \simeq 50 \Omega$, both located close to the nanowire to minimise parasitic inductances and capacitances. The nanowire acts as the inductor of a parallel RLC circuit whose impedance is :
  \begin{equation}
Z_{L}=\frac{1}{\frac{1}{R_L}+\frac{1}{jL_{tot} \omega}+jC_L\omega}
\end{equation}
The total inductance $L_{tot} = L_K+ L_G$ is the sum of a geometric term L$_G$ and a kinetic term L$_K$ due to the inertia of  Cooper pairs in the superconducting state. At the resonant frequency, the circuit impedance is purely real  $Z_L \simeq R_L \simeq 50 \Omega$. The microwave power is dissipated in the load circuit and a dip is expected in $|S_{11}|$. Figure \ref{Figure3} a shows a color-plot of S$_{11}$ as a function of frequency and temperature for the 200 nm wide meander. An absorption dip is evidenced (down to - 20 dB), at a resonant frequency f$_0$ which strongly decreases with temperature. We fitted the data for each temperature (see for instance the 30 K trace in the  inset Figure \ref{Figure3} a) with fixed values $C_{L}=24 pF$, the SMD components values. We therefore retrieve the inductance $L(T)$ which includes both $L_{tot}$ and a parasitic inductance, which is temperature independent. 

The geometric inductance for a wire of length $l$, width $w$ and thickness $e$ is $L_{G}=0.2 \cdot l \left(0.5 + (w+e)/3l + \ln \left( 2 \cdot l / \left(w+e \right)\right) \right)$ \cite{Yue}, while the kinetic one is $L_{K}=\mu_0\lambda(T)^2\cdot\left(l/we\right)$, where $\lambda$ is the London penetration depth of the superconductor. Within the two fluids model, and according to the Gorter Casimir (GC) expression, its temperature dependence is : $\lambda(T)^2=\lambda_0^2/\left(1-\left(T/T_{c}\right)^\alpha\right)$
where $\lambda_0$  is the zero-temperature penetration depth, $T_{c}$ the critical temperature and $\alpha$ a parameter whose value has been found to be 2 in the cuprates \cite{Rauch, ouanani, johansson, Wolf}.

In Figure \ref{Figure3} b, we plot $L_{tot}$ (symbols), determined from the resonance frequency, as a function of temperature for the 100 nm and 200 nm wide nanowires. We also show the best fits (solid lines) assuming $L_{tot}=L_{G}+L_{K}$ and the above formulas, and fixing $T_{c}$ as $T_{c0}$ according to DC measurements (see above), together with $\alpha=2$. The only free parameter is therefore $\lambda_{0}$ which is found to be 190 $\pm$ 8 nm and 181 $\pm$ 8 nm for the 100 nm and 200 nm wide wires respectively. These values are in very good agreement with previously reported ones in the literature \cite{Rauch, Wolf}.

In conclusion, we have presented a new reliable method to make \YBCO nanowires in a meander shape, based on high energy ion irradiation. Good superconducting characteristics are preserved for devices as narrow as 100 nm wide, over lengths up to $450\ \mu m$. Protected by a top \ceo layer and embedded in an insulating \YBCO matrix, the nanowires are robust, stable in time, and sustain thermal cycling. Using a resonant method, we have accurately measured their temperature dependent inductance - a key parameter for high speed operation - which perfectly agrees with the expected one. This work paves the way for High Tc based Superconducting Single Photon Detectors. 

The authors thank Yann Legall for ion irradiations, Jean-Claude Vill\'egier and M. Rosticher for fruitful discussions, T. Ditchi and J. Lucas for RF experiments support. This work has been supported by the french agencies ANR and DGA, by the Region Ile-de-France in the framework of DIM Nano-K and by the European Community (NQO  COST Action). \\
\end{document}